\documentclass{elsart5p}

\usepackage{amsmath} 
\newcommand{\plotwidth}{1.0}

\usepackage{epsfig}

\usepackage{amssymb}

\usepackage{lineno}

\begin{document}

\begin{frontmatter}



\title{Energy Loss Signals in the ALICE TRD}


\author{Xian-Guo Lu}

\address{Physikalisches Institut der Ruprecht-Karls-Universit\"{a}t, Heidelberg, Germany}

 for the ALICE Collaboration

\ead{Xianguo.Lu@cern.ch}

\begin{abstract}
We present the energy loss measurements with the ALICE TRD in the $\beta\gamma$ range 1--10$^{4}$, where
$\beta=v/c$  and $\gamma=1/\sqrt{1-\beta^2}$. The measurements are conducted in three different scenarios: 1)~with
pions and electrons from testbeams; 2)~with protons, pions and electrons in proton-proton collisions at center-of-mass energy 7 TeV; 3)~with muons detected in ALICE cosmic runs. In the testbeam and cosmic ray measurements, ionization energy loss (dE/dx) signal as well as ionization energy loss plus transition radiation~(dE/dx+TR) signal are measured.  With cosmic muons the onset of TR is observed.  Signals from TeV cosmic muons are consistent with those from GeV electrons in the other measurements. Numerical descriptions of the signal spectra and the~$\beta\gamma$-dependence of the most probable signals are also presented.

\end{abstract}

\begin{keyword}
Ionization energy loss \sep dE/dx \sep Transition radiation  \sep TR  \sep Cosmic muon
\PACS
\end{keyword}
\end{frontmatter}

\section{Introduction}
\label{sec:intro}

\begin{figure}[!h]
\begin{center}
\epsfig{file=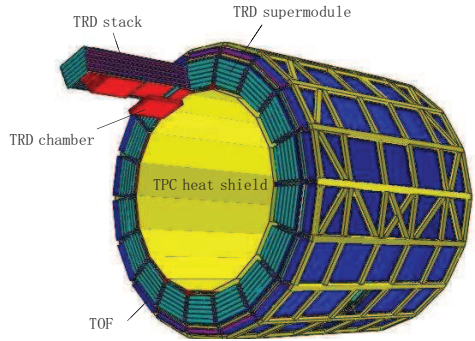, width=0.7\columnwidth}
\caption{Layout of the ALICE TRD.}\label{fig:trdgeo}
\end{center}
\end{figure}

The TRD~\cite{alitrd} of the ALICE experiment~\cite{alice} at the LHC is devoted to electron identification~\cite{ypach} and tracking of charged particles. It provides also Level-1 triggering on electrons and jets~\cite{jklei}. It is a cylindrical detector system located in radius between 2.9 and~3.7~meters from the beamline and segmented in 6 layers. It has a 2$\pi$ azimuthal coverage in 18 super-modules and a polar coverage between 45$^{\circ}$ and 135$^{\circ}$ in 5 stacks~(Fig.~\ref{fig:trdgeo}).

\begin{figure}[!h]
\begin{center}
\epsfig{file=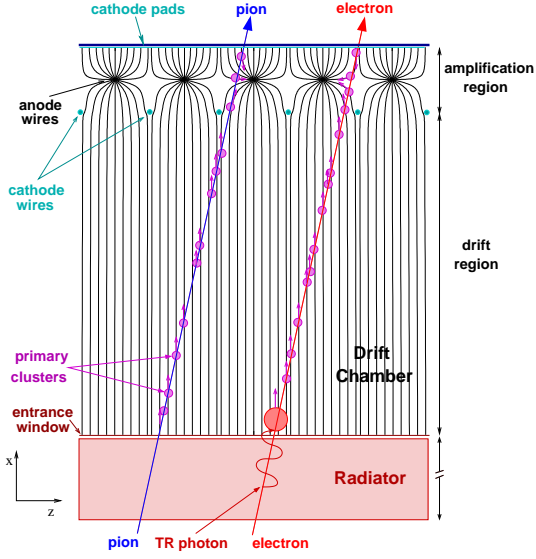, width=0.8\columnwidth}
\caption{Schematic cross-sectional view of a ALICE TRD chamber.}\label{fig:trdprin}
\end{center}
\end{figure}

Individual TRD chamber consists of a 4.8 cm thick layer of fibres/foam sandwich radiator and a drift chamber filled with Xe, CO$_{2}$ (15\%). The depths of the drift and amplification regions are 3 cm and 0.7 cm respectively. The induced  charges are readout by cathode pads which have a typical size of 0.7$\times$8.8 cm$^{2}$~(Fig.~\ref{fig:trdprin}) every 100 ns. A charged particle loses energy in primary collisions with gas atoms by ionization (dE/dx). Energetic ones with $\gamma$ above $10^3$ in addition emit   TR photons when passing through the radiator. The radial positions of the primary clusters are reconstructed from the drift time. 
 Therefore besides the energy loss measurement, the TRD is also used for momentum determination. The TRD signals presented in the following sections are  the integrated charge over all drift time corrected for the non-perpendicular incident angle.

\section{TRD Signals from Testbeam Measurement and Proton-Proton Collisions}
\label{sec:tbpp}

\begin{figure}[!h]
\begin{center}
\epsfig{file=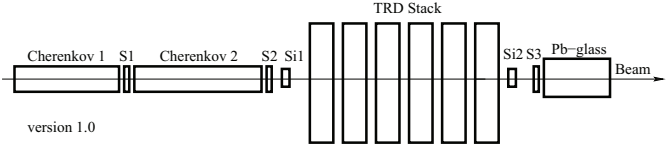, width=\plotwidth\columnwidth}
\caption{Beam line setup of TRD testbeam measurement~\cite{david}. Scintillator S1--3 are trigger detectors. Silicon strip detector Si1--2 locate the beams. Cherenkov counter 1--2 and lead glass calorimeter are used to identify pions and electrons. }\label{fig:tbset}
\end{center}
\end{figure}

The testbeam measurement was carried out at CERN's PS in 2004~\cite{tb2004} with secondary beams of pions and electrons of momenta from 1 to 10 GeV/c. The setup is shown in Fig.~\ref{fig:tbset}.
Prototype chambers without (with) radiators were used for the dE/dx (dE/dx+TR) measurement.

\begin{figure}[!h]
\begin{center}
\epsfig{file=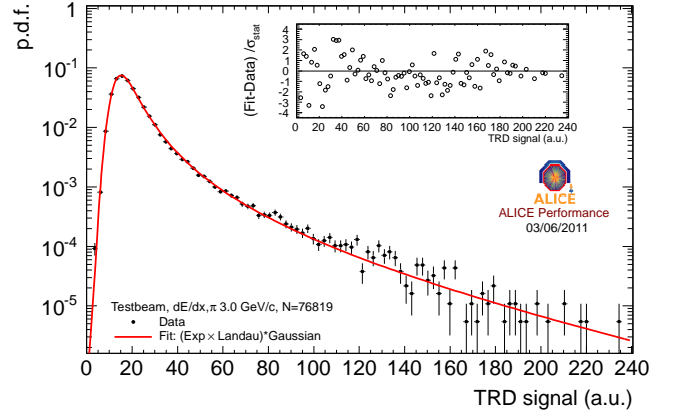, width=\plotwidth\columnwidth}
\epsfig{file=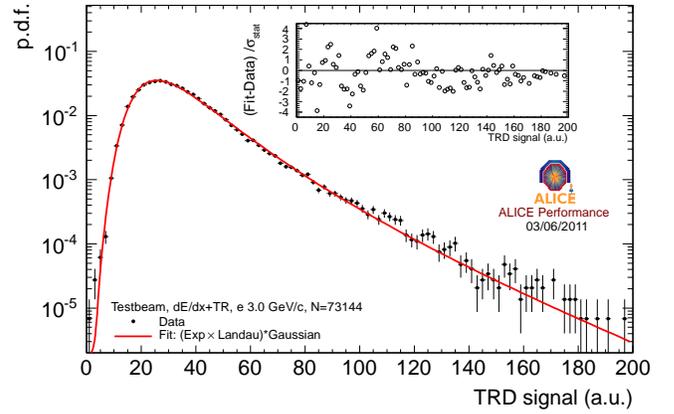, width=\plotwidth\columnwidth}
\caption{TRD signals from 3 GeV/c pions (\textit{upper}) and electrons~(\textit{lower}).}\label{fig:fit0}
\end{center}
\end{figure}

\begin{figure}[!h]
\begin{center}
\epsfig{file=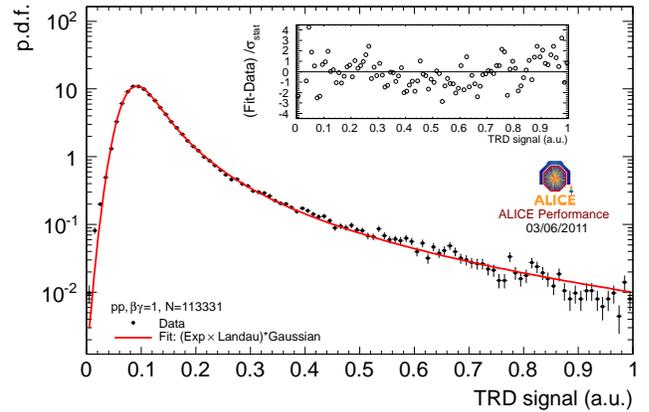, width=\plotwidth\columnwidth}
\caption{TRD signals measured at $\beta\gamma=1$ from identified protons in proton-proton collisions.}\label{fig:fitpp}
\end{center}
\end{figure}

\begin{figure}[!h]
\begin{center}
\epsfig{file=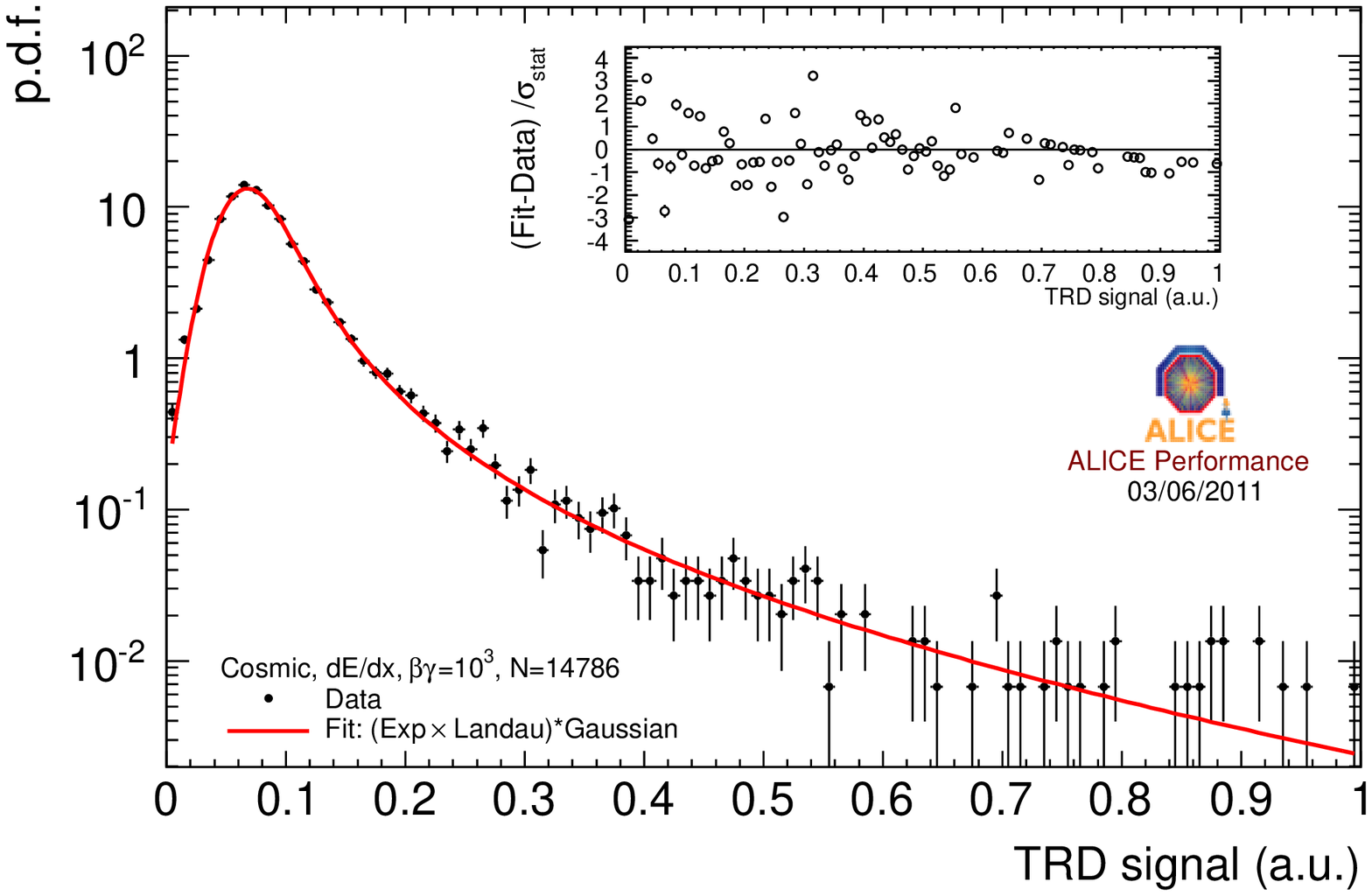, width=\plotwidth\columnwidth}
\epsfig{file=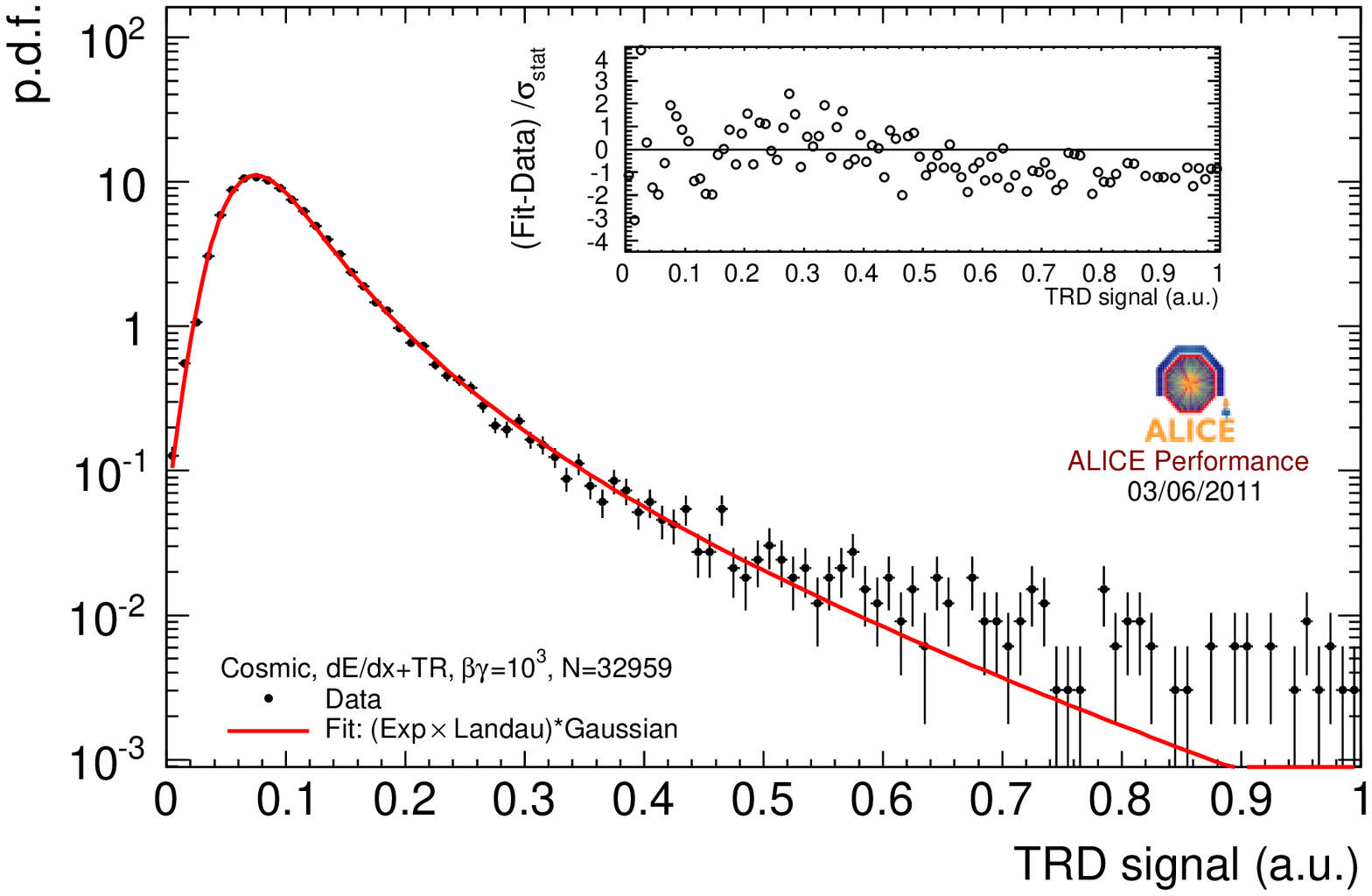, width=\plotwidth\columnwidth}
\caption{TRD signals measured at $\beta\gamma=10^3$ from cosmic runs. (\textit{Upper}:~dE/dx, \textit{lower}:~dE/dx+TR, see Section~\ref{sec:trdcosmic} for details.)}\label{fig:fitcos}
\end{center}
\end{figure}

In Figure~\ref{fig:fit0} the TRD signal distributions from 3~GeV/c testbeam pions and electrons are shown. Both  are fit with the following modified Landau-Gaussian convolution:
\begin{align}
(\textrm{Exponential}\times\textrm{Landau})*\textrm{Gaussian},
\end{align}
where the Landau distribution is weighted by an exponential damping
\begin{align}
\textrm{Landau}(x)\rightarrow e^{-\kappa x}\textrm{Landau}(x).
\end{align}
This function is used also for distributions from proton-proton collisions (Fig.~\ref{fig:fitpp}) and cosmic runs (Fig.~\ref{fig:fitcos}) to extract the most probable energy loss.

Since the end of March in 2010, ALICE has collected data from proton-proton collisions at center-of-mass energy 7 TeV. The TRD signals are measured for protons, pions and  electrons in minimum bias events. 
More details can be found in Ref.~\cite{mfasel}.


\section{TRD Signals from Cosmic Muons}
\label{sec:trdcosmic}

\begin{figure}[!h]
\begin{center}
\vspace{1cm}
\epsfig{file=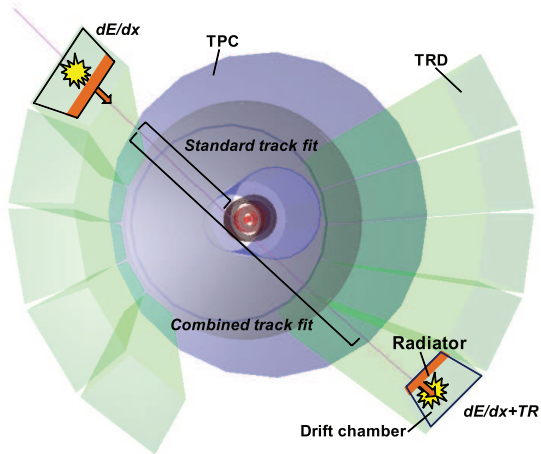, width=0.9\columnwidth}
\caption{One cosmic event in ALICE TPC and TRD (3D view along the beam pipe).}\label{fig:cosdisplay}
\end{center}
\end{figure}

The TRD energy loss measurements in the testbeam and proton-proton runs described in Section~\ref{sec:tbpp} do not cover the~$\beta\gamma$ range 10$^2$ -- 10$^3$ which can be filled in by cosmic muons. Because the ALICE detector is situated underground with 28 meters of material above, cosmic rays which leave long trajectories in the ALICE Time Projection Chamber (TPC)~\cite{alitpc} are predominantly muons. Because the TR photons are emitted in direction of the passage, they enter the drift section and are absorbed by the heavy gas when the particle traverses the radiator first. Contrarily,  only the dE/dx signal is measured when the particle  traverses  the drift section and then the radiator.  Due to the cylindrical placement of the  TRD layers, dE/dx signals are associated with the in-coming passages of the  cosmic muons and dE/dx+TR signals with the out-going ones~(Fig.~\ref{fig:cosdisplay}).

\begin{figure}[!h]
\begin{center}
\epsfig{file=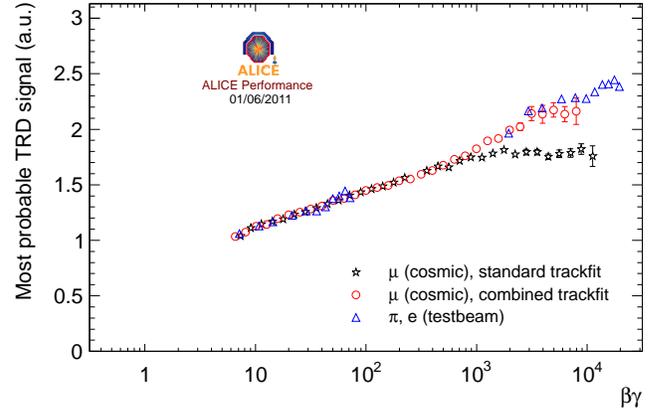, width=\plotwidth\columnwidth}
\caption{Most probable TRD dE/dx+TR signals in testbeam and cosmic ray measurements.}\label{fig:oldnew}
\end{center}
\end{figure}

In order to measure precisely the momenta of the muons, a specific track fitting  program was developed. Compared to the standard tracking which only fits half of the muon trajectory in the TPC, the combined track fit uses all $\mathrm{TPC}$ clusters~(Fig.~\ref{fig:cosdisplay}) and therefore achieves  a 10 times better momentum resolution: the $1/p_t$-resolution is 8.1$\times 10^{-4}$ c/GeV at momentum 1 TeV/c (integrated for all cosmic ray geometries in the TPC). Figure~\ref{fig:oldnew} shows the most probable TRD dE/dx+TR signals measured in the testbeam and cosmic runs. At $\beta\gamma$ above 10$^3$, the cosmic ray signal by the standard TPC tracking flattens as a result of the limited momentum resolution. With the improvement provided by the combined track fit, the cosmic-ray and  testbeam results are consistent up to $\beta\gamma=10^4$, beyond which the statistics is limited.

\begin{figure}[!h]
\begin{center}
\epsfig{file=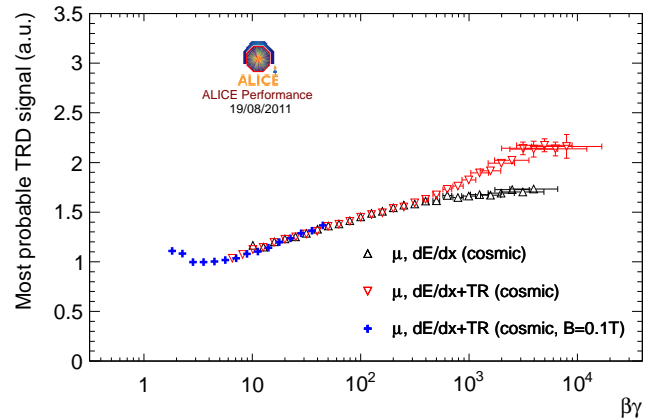, width=\plotwidth\columnwidth}
\caption{Most probable TRD signals from cosmic muons. Horizontal error bars are obtained from the estimated momentum resolution.}\label{fig:cosfull}
\end{center}
\end{figure}

The standard ALICE 0.5 T solenoidal magnetic field is too strong for sub-GeV muons to leave long trajectories in the TPC. In order to measure the muon minimum ionization signals in the TRD, cosmic runs with a magnetic field of 0.1~T were conducted. The momentum is determined by the  standard TPC-TRD tracking, which in addition to the standard TPC tracking also takes into account the space points in the TRD. This is the first ALICE running at 0.1~T.  The most probable TRD signals from cosmic muons are shown in Fig.~\ref{fig:cosfull}. By comparing the dE/dx and dE/dx+TR signals, the TR onset is observed at $\beta\gamma>700$.

\section{Results}

\begin{figure}[!h]
\begin{center}
\epsfig{file=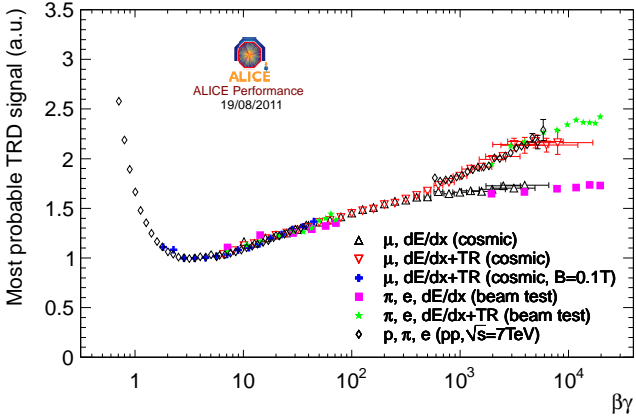, width=\plotwidth\columnwidth}
\caption{Most probable TRD signals from testbeam, proton-proton collisions and cosmic ray measurements.}\label{fig:dedxall}
\end{center}
\end{figure}

Combining the  testbeam, proton-proton collisions and cosmic ray measurements we obtained the TRD energy loss signals in the $\beta\gamma$ range 1 -- 10$^4$~(Fig.~\ref{fig:dedxall}). Results from different measurements are consistent.

\begin{figure}[!h]
\begin{center}
\epsfig{file=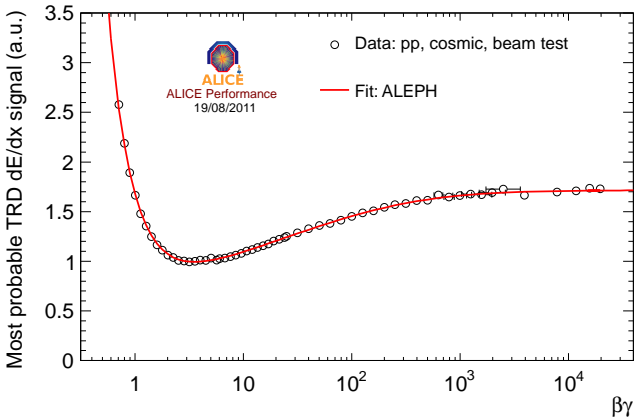, width=\plotwidth\columnwidth}
\caption{Compilation of TRD dE/dx data from testbeam, proton-proton collisions and cosmic ray measurements.}\label{fig:fitdedx}
\end{center}
\end{figure}

The most probable (mp) TRD dE/dx signal in unit of minimum ionization is well described by the ALEPH~\cite{aleph, aleph2} parameterization~(Fig.~\ref{fig:fitdedx}):
\begin{align}
\frac{\mathrm{d}E}{\mathrm{d}x}_\mathrm{mp}=0.20\times\frac{4.4-\beta^{2.26}-\ln\left[0.004+\frac{1}{\left(\beta\gamma\right)^{0.95}}\right]}{\beta^{2.26}}.
\end{align}
From the fit result two numbers of practical interest can be deduced:
\begin{enumerate}
\item $\beta\gamma$ of minimum ionizing particle: 3.5,
\item dE/dx in the relativistic limit: 1.7 times minimum ionization.
\end{enumerate}

\begin{figure}[!h]
\begin{center}
\epsfig{file=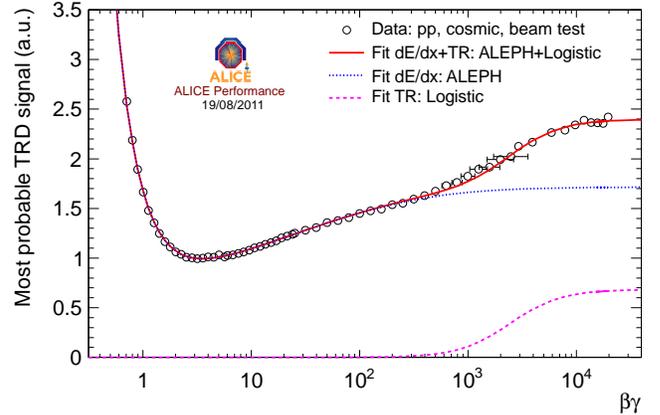, width=\plotwidth\columnwidth}
\caption{Compilation of TRD dE/dx+TR data from testbeam, proton-proton collisions and cosmic ray measurements. The ALEPH component of the fit is fixed with the dE/dx results in Fig.~\ref{fig:fitdedx}.}\label{fig:fitdedxtr}
\end{center}
\end{figure}

Fixing the dE/dx contribution according to the ALEPH parameterization, the most probable dE/dx+TR signal in unit of minimum ionization is well described by including an additional Logistic function~(Fig.~\ref{fig:fitdedxtr}):
\begin{align}
\mathrm{TR_{mp}}=\frac{0.706}{1+\exp^{-1.85\times(\ln\gamma-7.80)}}.
\end{align}
The following practical  TR properties can be deduced:
\begin{enumerate}
\item saturated TR  yield in the relativistic limit: 0.7 times minimum ionization,
\item $\beta\gamma$ for half saturation: $2.4\times10^3$.
\end{enumerate}

\section{Summary}
\label{sec:summary}

We have presented an overview of the energy loss $\mathrm{signals}$ in the
ALICE TRD, measured with prototypes in a testbeam and with the actual detector in proton-proton collisions and cosmic runs in the ALICE setup at the LHC. In the cosmic ray measurement we exploit the geometry of the
detector to measure separately the dE/dx and dE/dx+TR signals in the TRD. In addition, a combined track fit within the ALICE TPC was developed to exploit the full length of the track, leading to a $1/p_t$-resolution of  8.1$\times 10^{-4}$ c/GeV at momentum 1 TeV/c.
The TR from TeV cosmic muons is unambiguously observed.
We have shown numerical descriptions of the signal spectra and of the dependence of the most probable energy loss on $\beta\gamma$, which will allow to establish reference distributions for
particle identification with TRD over a broad momentum range.




\end{document}